\documentstyle[aps,epsf,twocolumn]{revtex}
\newcommand{\NEW}[1]{#1}
\newcommand{\OLD}[1]{}

\newcommand{\be}{\begin{equation}}
\newcommand{\ch}{{\rm ch}}
\renewcommand{\vec}[1]{{\bf #1}}
\newcommand{\ee}{\end{equation}}
\newcommand{\bea}{\begin{eqnarray}}
\newcommand{\eea}{\end{eqnarray}}
\newcommand{\vx}{{\bf x}}
\newcommand{\vp}{{\bf p}}
\newcommand{\vv}{{\bf v}}

\begin{document}
\title{New solutions to covariant nonequilibrium dynamics}

\draft
\author{D\'enes Moln\'ar and Miklos Gyulassy}

\address{Physics Department, Columbia University,
         538 W. 120th Street, New York, NY 10027}

\date{%
\OLD{%
Received 22 May 2000
}
\NEW{%
Original submission 22 May 2000, revised version 13 July 2000%
}
}

\maketitle

\begin{abstract}
New solutions of 3+1D covariant kinetic theory are presented
for nuclear collisions in the energy domain $E_{cm}\sim 200$ $A$GeV.
They are obtained using MPC,
a new Monte-Carlo 
parton transport technique that employs very high parton subdivision
that is necessary to preserve covariance.
The transport results are compared with ideal hydrodynamics solutions%
\OLD{%
 obtained employing Cooper-Frye isotherm freeze-out prescriptions%
}
.
We show that the transport evolution differs significantly
from hydrodynamics.
In addition, 
\OLD{%
the
}
\NEW{%
we compare the transport freeze-out distributions
to those obtained from ideal hydrodynamics
with the Cooper-Frye isotherm freeze-out prescription.
The transport
}
freeze-out four-volume is shown to be
sensitive to the reaction rates and
deviates from both timelike and spacelike freeze-out 3D
hypersurfaces commonly assumed.
In particular, we find that there does not exist a universal
freeze-out temperature. Finally,
the  transverse momentum distributions are found to
deviate by up to an order of  magnitude
from (Cooper-Frye frozen) hydrodynamics
for a wide range of possible initial conditions
and reaction rates at RHIC energies.
\end{abstract}


\section{Introduction and Conclusions}
\label{Section:intro}

A theoretical framework to study nonequilibrium dynamics is provided
by Boltzmann transport theory.  The dynamical variables of this theory
are the Lorentz-covariant, one-particle phase space distributions
$f_i(x, p)$; while the dynamics is governed by transition
probabilities $W_{c\to c^\prime}$, which are Lorentz-covariant
functions of the particle momenta.
The theory, while not exact, is rather general.
First, it is not restricted to particular particle types.
The particles could be partons, hadrons, or molecules.
Second,
the reaction rates that specify the dynamics
are also unrestricted in their origin.
For example,
the rates could emerge from an effective quantum field theory
or Newtonian mechanics.
The primary limitations of the theory
are the neglect of dynamical correlations
(one-body truncation of the formal BBGKY hierarchy\cite{degroot})
and the inability,
without additional classical fields,
to model phase transition dynamics.

We consider here
the simplest form of Lorentz-covariant Boltzmann transport theory
in which the on-shell phase space density $f(x,\vp)$,
evolves with an elastic $2\to 2$ rate as\cite{Yang,Zhang:1998ej,MPC}:
\bea
p_1^\mu \partial_\mu f_1 &=&\int\limits_2\!\!\!\!
\int\limits_3\!\!\!\!
\int\limits_4\!\!
\left(
f_3 f_4 - f_1 f_2
\right)
W_{12\to 34} \delta^4(p_1+p_2-p_3-p_4)
\nonumber \\
&&
+  \, S(x, \vp_1) .
\label{Eq:Boltzmann_22}
\eea
Here $W$ is the square of the scattering matrix element,
the integrals are shorthands
for $\int\limits_i \equiv \int \frac{g\ d^3 p_i}{(2\pi)^3 E_i}$,
where $g$ is the number of internal degrees of freedom,
while $f_j \equiv f(x, \vp_j)$.
The initial conditions are specified by the source function $S(x,\vp)$,
which we discuss later in Section \ref{Section:transport_theory}.
For our applications below,
we  interpret  $f(x,\vp)$ as describing
an ultrarelativistic massless gluon gas 
with $g=16$ (8 colors, 2 helicities).

Recall several important properties of Eq. (\ref{Eq:Boltzmann_22}).
First, the particle number current and the energy-momentum tensor 
are given by
\be
N^\mu(x) \equiv \int \frac{d^3p}{(2\pi)^3 E} p^\mu f(x,\vec p)
\ee
and
\be
T^{\mu\nu}(x) \equiv \int \frac{d^3p}{(2\pi)^3 E} p^\mu p^\nu f(x,\vec p).
\ee
With these definitions,
particle number and energy-momentum conservation
{\em follow} from Eq. (\ref{Eq:Boltzmann_22})
(when the source term $S(x,\vp)= 0$).
Second, there is a  class of fixed points, called {\em global equilibria},
which are  phase space densities of the form
 \be
f(x,\vec p)=\frac{g}{(2\pi)^3}
\,\exp\!\left[\frac{\mu-p_\mu u^\mu}{T}\right],
\label{Eq:global_equil_distribution}
\ee
where $u_\mu$ is a constant four-vector that specifies a global flow velocity,
while $T$ and $\mu$ correspond
to the constant temperature and chemical potential.
Furthermore,
the {\em H-theorem}\cite{degroot} states
that the Boltzmann transport equation
drives the system towards global equilibrium.

Despite its relatively simple form, the Boltzmann equation is
nonlinear with very few known analytic solutions.  Until recently,
progress to obtain even numerical solutions has been hampered by its
numerical complexity.  The rapid increase in computational power has
finally made it possible to break through this barrier.
For nuclear collision applications,
new numerical algorithms are being developed,
tested,
and made available via the World Wide Web
under a new Open Standard for Codes and Routines (OSCAR)\cite{OSCAR}.
The present work is a further step in that development.

For nuclear collisions at SPS energies ($\sqrt{s}\le 20$ $A$GeV),
numerical solutions of hadronic transport models have been available
for some time\cite{Urqmd:1999xi}.
However,
for higher collider energies,
the emergence of massless partonic degrees of freedom
creates the technical challenge of how to retain Lorentz covariance.
In this paper we present results based on a new numerical technique,
MPC~0.1.2~\cite{MPC}, that provides reliable solutions in this
ultrarelativistic regime.

The difficulty of the analytic treatment
of the Boltzmann transport equation
has forced workers in the past to make strong simplifying assumptions.
A common simplification has been
to ignore the general nonequilibrium problem
and to postulate  that {\em local equilibrium} is maintained at all times.
In the framework of the Boltzmann transport theory,
this choice corresponds to substituting
a local equilibrium ansatz in place of the fixed point  global equilibria.
Allowing $(u_\mu, T, \mu)$ to vary with the coordinate $x^\mu$,
this ansatz corresponds to
\be
f(x,\vec p)=\frac{g}{(2\pi)^3}
\,\exp\!\left[\frac{\mu(x)-p_\mu u^\mu(x)}{T(x)}\right].
\label{Eq:local_equil_distribution}
\ee It is however well known that local equilibrium is {\em not} a
solution of Eq. (\ref{Eq:Boltzmann_22}).
 The nonlinear collision term vanishes in local equilibrium,
but $p^\mu \partial_\mu f \not = 0$ in general. 
Only in the limit when all the rates go to infinity,
i.e., when the mean free path goes to zero, can the solution approach
local equilibrium.

A covariant, dynamical theory can nevertheless be postulated based on
the assumption of local equilibrium.
That is relativistic hydrodynamics\cite{La53},
which is widely used in heavy-ion
physics\cite{Cs94,ECT97} to calculate observables.
Assuming an equilibrium initial condition
specified on a hypersurface $\sigma_{in}^\mu(x)$,
the local energy momentum and baryon number conservation laws 
\be \partial^\mu T_{\mu\nu} = 0,
\qquad \partial^\mu N_{B,\mu} = 0
\label{Eq:cons_laws}
\ee
reduce to (Euler) hydrodynamical equations under the assumptions
that local chemical and thermal equilibrium are maintained
and that dissipation (viscosity and thermal conductivity) can be neglected.
In that case,
the energy-momentum tensor and baryon current can be expressed
as $T_{\mu\nu}=u_\mu u_\nu(e+p)-g_{\mu\nu}p$
and $N_{B,\mu} =u_\mu n(x)$
in terms of the local flow velocity $u_\mu(x)$,
local pressure $p(x)$,
local energy density $e(x)$,
and local proper density $n(x)$.
The equations form a closed system if,
in addition,
the equation of state, $e(p,n_B)$,
is specified.

It is clear that the idealization of local equilibrium may  apply,
if at all,
only in the interior of the reaction volume,
where the local mean free path
 $\lambda(x)=1/(\sigma n(x))$
may remain  small for a while
as compared to the characteristic dimensions and gradients of the system, 
$L_\mu(x) \sim|\partial_\mu \log e(x)|^{-1}$.
However,
these assumptions are marginal
for conditions encountered in heavy ion collisions
and certainly break down near the surface region
and throughout the freeze-out phase.
Due to longitudinal expansion the density decreases as $1/\tau$
until $\tau\sim \sqrt{3}R$ when 3-dimensional expansion
rapidly increases $\lambda$ beyond $L$.

For small departures from local equilibrium,
corrections to the ideal Euler hydrodynamic evolution can be calculated
by taking the $T^{\mu\nu}$ and $N_B^\mu$ moments
of an underlying kinetic theory.
To first order in $\lambda/L$,
the equations reduce to the Navier-Stokes equations.
The solutions to Navier-Stokes depend therefore
not only on the equation of state,
but also on the bulk and shear viscosity
and thermal conductivity transport coefficients of the medium.
While such an approach has proven useful
in nonrelativistic problems and in special relativistic geometries,
severe problems of instability and acausality appear
when extended into the ultrarelativistic domain\cite{Olson:1990eu}.

The newly formulated, covariant, parton kinetic theory technique,
MPC\cite{MPC}, allows us to compute the highly dissipative evolution
during the densest partonic phase of the reaction in a covariant
manner as well as investigate the final freeze-out dynamics.  MPC is
an extension of Zhang's covariant parton cascade algorithm,
ZPC\cite{Zhang:1998ej}.
Both MPC and ZPC have been
extensively tested \cite{Gyulassy:1997ib,Zhang:1998tj} 
and compared to analytic transport solutions
and covariant Euler and Navier-Stokes dynamics in 1+1D geometry.
A critical new element of both these algorithms
is the parton subdivision technique proposed by Pang\cite{Yang,Zhang:1998tj}.
As shown in detail in Section \ref{Section:subdivision},
rather high subdivision $\sim 100$ is needed
to preserve Lorentz covariance numerically for massless parton evolution.

Extensions of MPC 
to include inelastic $2\leftrightarrow 3$ partonic
processes\cite{molnar:QM99} are under development,
but in this paper we use MPC
in the pure elastic parton interactions mode as in ZPC\cite{OSCAR}.
 
The aim of this work is to calculate the sensitivity
of the evolution and freeze-out
of an ultrarelativistic (massless) parton system
to the transport rates and initial conditions
expected in Au+Au reactions at RHIC energies ($\sqrt{s}\sim 200$ AGeV). 
We compare the results  to the (OSCAR compliant)
relativistic Euler hydrodynamic code
developed by Rischke and Dumitru \cite{OSCAR,rischdumi}.
An ideal gas ($e=3p$) equation of state is used in that analysis.
This work extends Refs. \cite{Gyulassy:1997ib,Zhang:1999bd},
focusing on the freeze-out problem in 3+1D Bjorken expansion.
It is less ambitious than  for example Ref. \cite{Zhang:1999bd}
by limiting the study to massless partons.
This avoids introducing yet further complications
due to dissipative hadronization and hadronic transport effects.
The  initial conditions
are taken from the HIJING multiple mini jet generator%
\cite{OSCAR,Gyulassy:1994ew}.

\begin{figure}[hp]
\center
\leavevmode
\epsfysize=10cm
\epsfbox{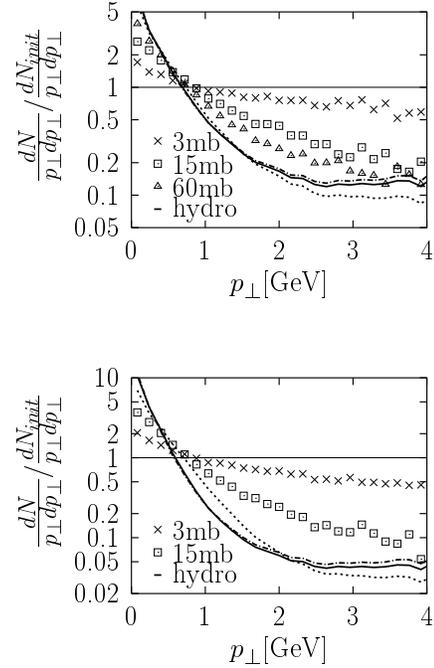}
\caption{
\footnotesize
The top figure shows the freeze-out $p_\perp$-distributions
relative to the {\em initial} ($T_0=500$ MeV)
$p_\perp$-distribution
from an initial radius $R_0=2$ fm
for the cascade with $\sigma=3$, 15, 60 mb,
and for ideal hydrodynamics with Cooper-Frye freeze-out
with freeze-out temperatures $T_f=$100 MeV (dashed-dotted line),
130 MeV (thick solid line), and 200 MeV (dotted line).
The bottom figure shows the same but for an initial radius $R_0=6$ fm
with cross sections 3 and 15 mb.
}
\label{Figure:2}
\label{Figure:fo_pt_ratio}
\end{figure}

\OLD{%
One of the main results of this work
is shown in Fig. \ref{Figure:fo_pt_ratio}.
To test whether the hydrodynamic assumption of local equilibrium
is justified by transport theory,
we have followed the evolution
both for the Boltzmann equation (\ref{Eq:Boltzmann_22})
and Euler hydrodynamics
from the {\em same} RHIC initial condition from \cite{Gyulassy:1994ew}.
(See Section \ref{Section:results} for details of the simulations.)
Figure \ref{Figure:fo_pt_ratio} shows
the final, experimentally observable $p_\perp$-distributions
divided by the thermal initial distribution.
We find that  there is an order of magnitude difference
between transport and hydrodynamic solutions
at {\em both} low and at high $p_\perp$
for realistic ($\sim$ few mb) gluonic cross sections.
The difference is still a factor of two to three
even for a physically extreme, 15 mb, cross section.
This latter curve is equivalent%
to a solution for  3 mb cross section
with five times higher\cite{Eskola:2000fc} initial density
than expected with HIJING.
In addition,
increasing the radius of the initial Bjorken cylinder
from 2 to 6 fm does not reduce the  discrepancy
between the covariant transport solutions and those of hydrodynamics.
Therefore, large deviations from local equilibrium and hence hydrodynamics
are to be expected for possible initial conditions at RHIC.
}

\NEW{%
One of the main results of this work
is shown in Fig. \ref{Figure:et_evolution}.
To test whether ideal hydrodynamics is an adequate approximation
of the parton transport equation (\ref{Eq:Boltzmann_22}),
we have followed the evolution
both for the Boltzmann equation (\ref{Eq:Boltzmann_22})
and Euler hydrodynamics
from the {\em same} RHIC initial condition from \cite{Gyulassy:1994ew}.
(See Section \ref{Section:results} for details of the simulations.)
Figure \ref{Figure:et_evolution} shows the evolution of the transverse energy
$dE_t/dy$ at midrapidity.
There is a large difference
between the transport and hydrodynamic solutions
in both the 1+1 and 3+1 dimensional case,
even for a physically extreme, 15mb,
cross section.
Note, that these 15mb curves are equivalent%
\footnote{
The equivalence is due to the scaling property of Eq. (\ref{Eq:Boltzmann_22})
explained in Section \ref{Section:subdivision}.
}
to a solution for  3 mb cross section
with five times higher\cite{Eskola:2000fc} initial density
than expected with HIJING.
Therefore,
large deviations from ideal hydrodynamics
are to be expected for possible initial conditions at RHIC.
In addition,
this conclusion is independent of the initial system size
(see explanation in Subsection \ref{SubSection:evolution_comparison}).%
}

This conclusion reinforces
the results of  Ref. \cite{Gyulassy:1997ib},
where it was  shown
that large deviations even from the Navier-Stokes evolution
in 1+1D Bjorken expansion are expected
for initial densities up to four times higher
than predicted by the HIJING model\cite{Gyulassy:1994ew}.

\NEW{%
A second main result of this work
is shown in Fig. \ref{Figure:fo_pt_ratio}.
We tested whether
the widely used Cooper-Frye freeze-out prescription
could ``correct'' final observables
for the neglect of the early breakdown of ideal hydrodynamics.
Figure \ref{Figure:fo_pt_ratio} shows
the final, experimentally observable $p_\perp$-distributions
divided by the thermal initial distribution.
We find that  there is an order of magnitude difference
between transport and hydrodynamic solutions
at {\em both} low and at high $p_\perp$
for realistic ($\sim$ few mb) gluonic cross sections.
The difference is still a factor of two to three
even for a physically extreme, 15 mb, cross section.
In addition,
increasing the radius of the initial Bjorken cylinder
from 2 to 6 fm does not reduce the  discrepancy
between the covariant transport solutions
and those of Cooper-Frye frozen hydrodynamics.
To get closer to the transport $p_\perp$-distributions,
one would have to choose a freeze-out temperature
much above the commonly assumed $100-150$~MeV range.
}

\OLD{%
A second main result
}
\NEW{%
The last main result
}
of this work is illustrated 
in Fig. \ref{Figure:fo_spacetime_6fm}.
This  shows that high hydrodynamic freeze-out temperatures
that would be needed to ``fit'' the transport solutions are,
however, inconsistent with the space-time freeze-out distributions
of covariant transport theory.
Unlike the ``sharp'' space-time freeze-out particle distributions
commonly assumed using the Cooper-Frye freeze-out prescription, the
transport theory freeze-out volume is four-dimensional.  Particles
freeze out over a large four-volume that forms a wedgelike freeze-out
region in the $\tau-R$ plane.  This freeze-out distribution depends
strongly on the microscopic reaction rates (higher rates lead to a
later freeze-out).  It is {\em not} possible to tune the Cooper-Frye
freeze-out temperature to reproduce the cascade freeze-out
distributions.  Though one can arrange that the Cooper-Frye freeze-out
curve follows more-or-less the ridge,
the transport distribution {\em along} that ridge
is not correctly reproduced by
\NEW{%
Cooper-Frye frozen
}
hydrodynamics.
In particular, hydrodynamic freeze-out surfaces with a timelike section
result in unphysical spikes in the $dN/d\tau$ distribution which are
not present in the transport theory calculations.
These spikes arise when the freeze-out temperature is such that the 
interior of the system freezes out due to longitudinal Bjorken
expansion (see Subsection \ref{SubSubSection:coordinate_space} for
further discussion).

In summary, our results show that for a rather wide range of initial
conditions at RHIC energies, the evolution of the system deviates
strongly from 
\OLD{%
local equilibrium 
}
\NEW{%
Eulerian hydrodynamics
}
throughout the 3+1D evolution.
\NEW{%
It is not possible to mimic the observables from the nonequilibrium evolution
by simply applying the isotherm Cooper-Frye freeze-out prescription
to ideal hydrodynamics.
}
The
space-time four-volume of freeze-out,
even for the largest ($R\sim 6$ fm) nuclei,
does not resemble a timelike surface.  In addition, the
observable transverse momentum spectra are very sensitive to the
microscopic reaction rates.
These results indicate that while
\NEW{%
ideal
}
hydrodynamics is a useful model to explore possible collective
dynamics in nuclear collisions, the interpretation of experimental
observables must take into account the finite transition probabilities
$W_{\{i\}\to\{j\}}$ that govern the nonequilibrium evolution.
Fortunately,
numerical techniques such as MPC and ZPC
are now readily available\cite{OSCAR}.
Experimentally,
the $A$, $E_{cm}$, and multiplicity dependence
of the observables provides the best way
to measure these effective reaction rates
in the ultradense matter formed in nuclear collisions.

\section{Covariant Parton Transport Theory}
\label{Section:transport_theory}

Equation (\ref{Eq:Boltzmann_22}) is the simplest form
of classical Lorentz-covariant Boltzmann transport theory.
In principle,
the transport equation could be extended for bosons
with the substitution $f_1 f_2 \to f_1 f_2 (1+f_3) (1+f_4)$
and a similar one for $f_3 f_4$
(where we used the short-hand $f_i\equiv f(x,{\vp}_i)$).
In practice, no covariant algorithm yet exists
to handle such nonlinearities.
We therefore limit our study to quadratic dependence of the collision
integral on $f$.

The elastic gluon scattering matrix elements in dense parton systems
are typically of the Debye-screened form: 
$d\sigma/dq^2\approx (9\pi \alpha_s^2/2)/(q^2+\mu^2)^2$, 
which favors small angle scattering\cite{Gyulassy:1997ib}.
However, the relevant transport cross section is
$\sigma_t = \int d\sigma \sin^2\theta_{cm}\approx (9\pi\alpha^2/2
s)\log(s/4\mu^2)$,
where $s\approx 17 T^2$.
In order to maximize the equilibration rate for a fixed cross section,
we take here an {\em isotropic} differential cross section
in the center-of-mass frame instead.
We further assume  an energy-independent cross section
with a threshold specified by $\mu^2$,
i.e., our solutions therefore
correspond to the microscopic dynamics specified by
the following idealized  model 
\be
d\sigma = \Theta(s^2-\mu^2) \frac{\sigma_0}{4\pi} d\Omega .
\label{Eq:cross_section}
\ee
The transport cross section is $2\sigma_0/3$ in this case.

It is important to emphasize that while the cross section suggests a
geometrical picture of action over finite distances, we use
Eq. (\ref{Eq:cross_section}) only as a convenient parametrization to
describe the effective {\em local} transition probability, $W$.  In
the present study this is simply modeled as
$dW/d\Omega=s\;d\sigma/d\Omega$.  The particle subdivision technique
(see next Section) needed to recover covariance removes all notion of
nonlocality in this approach, just like in hydrodynamics.
Thus, the cross sections, e.g., 60 mb, used in the present study to simulate
rapid local changes of the phase space density in no way imply that
distances bigger than $1$ fm play any role.

With the above cutoff $\mu$,
freeze-out of a test particle can arise in two different ways:
either the system becomes too {\em dilute},
i.e., $1/n\sigma_0 >>L$,
or the system {\em cools} down
and the threshold suppresses further interactions.
By construction,
the possibility for the latter case occurs along an isotherm,
$T_f\approx \mu/\sqrt{17}$.
With Eq. (\ref{Eq:cross_section}),
we can therefore study the influence of dissipative phenomena
by varying the two scales $\sigma_0$ and $\mu^2$.
The evolution was performed
with $\sigma_0=3$, 15, 30, 60, 121 mb and $\mu=0$, 0.1, 0.5 GeV.

The initial condition was taken to be
a longitudinally boost invariant Bjorken cylinder
in local thermal and chemical equilibrium
at temperature $T(\tau_0)=500$ MeV at proper time $\tau_0=0.1$~fm/$c$
as by fitting the gluon mini-jet transverse momentum spectrum
predicted by HIJING\cite{Gyulassy:1994ew}. 
In order to compare to hydrodynamics,
we assume that the transverse density distribution is uniform
up to a radius $R_0=2$, 4, 6, or 8 fm. 
The  pseudo-rapidity $\eta\equiv 1/2 \log((t+z)/(t-z))$ distribution
was taken as uniform between $|\eta| < 5$.
Since we want to compare
to chemically and thermally equilibrated hydrodynamics,%
\footnote{
Technically,
MPC was run with an out-of-chemical-equilibrium initial gluon density
$n_{\eta,0} =4$~fm$^{-2}$
as obtained via HIJING including final state radiation
and with cross sections $\sigma_0=2$, 10, 20, 40, and 80 mb.
As explained in Section \ref{Section:subdivision},
because of the scaling property of the solutions of
the transport equation, the solutions for the chemical equilibrium
initial condition are identical when the cross section is rescaled
by a factor $l=2.6505/4 \approx 1/1.509$.
}%
the equilibrium initial gluon density was taken for this $T(\tau_0)$ to be
\be
n_{\eta,0} \equiv \left.\frac{dN}{d\eta d^2x_\perp}\right|_{\tau_0}
= \frac{g}{\pi^2}T^3\tau_0 \approx 2.65\ {\rm fm}^{-2} .
\ee
Evolutions from different initial densities can be obtained
by varying the cross section only
and using the scaling property explained in the next Section.

\section{Parton Subdivision and Scaling of Solutions}
\label{Section:subdivision}

We utilize the parton cascade method
to solve the Boltzmann transport equation (\ref{Eq:Boltzmann_22}).
A critical drawback of all cascade algorithms is
that they inevitably lead to numerical artifacts
because they violate Lorentz covariance.
This occurs because particle interactions are assumed to occur
whenever the distance of closest approach (in the relative c.m.)
is $d<\sqrt{\sigma_0/\pi}$, which corresponds to action at a distance.
To recover the {\em local} character of equation (\ref{Eq:Boltzmann_22})
and hence Lorentz covariance,
it is essential to use
the parton subdivision technique\cite{Yang,Zhang:1998ej}.
This is based  on the covariance of Eq. (\ref{Eq:Boltzmann_22})
under the transformation 
\be f\to f'\equiv l\, f, \quad W\to W'\equiv W/l \quad
(\sigma\to \sigma' = \sigma/ l).
\label{Eq:particle_subdivision}
\ee
As shown in Ref. \cite{Zhang:1998tj},
the magnitude of numerical artifacts
is governed by the diluteness of the system
$\sqrt{\sigma}/\lambda_{MFP}$,
that scales with $1/\sqrt{l}$.
Lorentz violation therefore formally vanishes in the $l\to \infty$ limit.

\subsection{Convergence with subdivision}

\begin{figure}[hp]
\center
\leavevmode
\epsfysize=12cm
\epsfbox{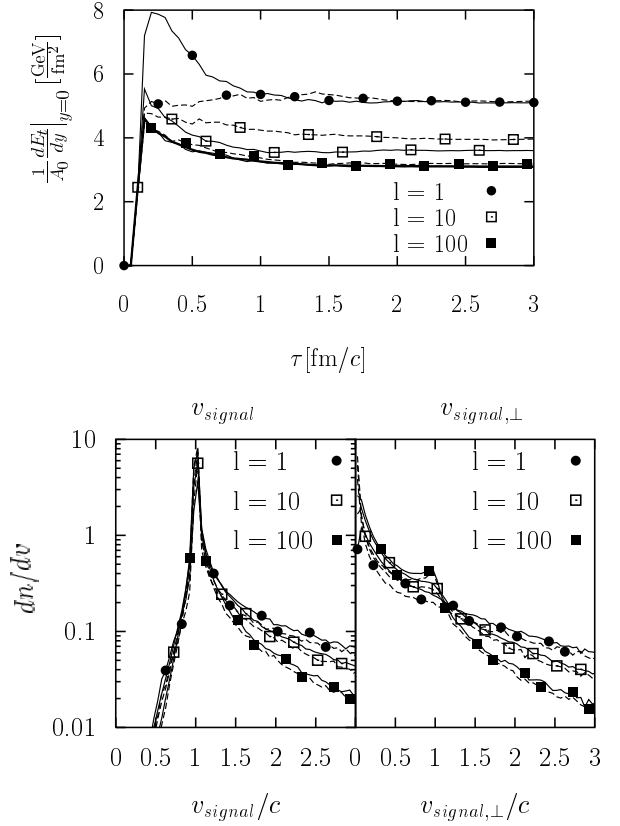}
\caption{
\footnotesize
The top figure shows the $E_t$ evolution at midrapidity (solid lines)
for subdivision factors $l=1$, 10, 100
from the initial condition $R_0 = 2$ fm,
$\sigma = 10$ mb,
$n_{\eta,0}=4$~fm$^{-2}$,
$T_0 =\mu = 0.5$ GeV,
$\tau_0 = 0.1\ {\rm fm}/c$.
The dashed curves are for the same initial condition
but all particles were boosted longitudinally by 3 units of rapidity
and $dE_t/dy$ was computed at $y=3$.
Though not visible, the curve for $l=400$ 
falls  on top of the $l=100$ curve.
The bottom plots show the normalized 
distribution of the signal propagation velocity
averaged over the full evolution.
}
\label{Figure:Et_and_signal_velocity}
\label{Figure:4}
\end{figure}

Figure \ref{Figure:Et_and_signal_velocity} illustrates
the severeness of the cascade numerical artifacts 
in the case of insufficient particle subdivision.
The top plot in Fig. \ref{Figure:Et_and_signal_velocity} shows
that the parton cascade solution for the evolution
 of the transverse energy per unit rapidity
does not converge
until the  subdivision factor reaches  $l\sim 100$.
The lack of covariance can be seen in the difference between the solutions
in frames separated by 3 units of rapidity.
The very fact that the cascade evolution is different
for different particle subdivisions means
that the subdivision covariance (\ref{Eq:particle_subdivision}) is itself
 violated
by the cascade algorithm.
Nevertheless,
both Lorentz and subdivision covariance are recovered
when $l$ is sufficiently large.

The large overshoot in the $dE_t/dy$ evolution
is a result of the superluminal signal propagation speed
inherent to the cascade algorithm.
A cascade particle can influence almost {\em instantaneously} 
another cascade particle that is within the interaction range
$r_\sigma\equiv \sqrt{\sigma / \pi}$.
In a very dense system,  a ``chain''
of almost instantaneous interactions can occur causing
long range superluminal artifacts.

As a measure of the signal propagation speed
in a nonlocal collision in the cascade we define
\be
\vec v_{s}
\equiv \frac{\vx_{partner}(t_{collision})
        - \vx_{particle}(t_{last\ collision})
       }
       {t_{collision}-t_{last\ collision}}.
\label{Def:signal_velocity}
\ee
Analytically, 
the deviation of the signal propagation speed
from the speed up light can be roughly approximated by
\be
\Delta v_s = \frac{r}{t},
\ee
where $t$ is the time between the collision and the previous collision,
while $r$ is the distance between the colliding particles
at the time of the collision ($r < r_\sigma$).
This is a pessimistic estimate
that maximizes the deviations.
Assuming that subsequent collisions are uncorrelated,
$t$ follows a Poisson distribution%
\footnote{
The scaling (\ref{Eq:particle_subdivision})
leaves the mean free path $\lambda$ invariant.
}
\be
P(t) \equiv \frac{dn}{dt} = \frac{1}{\lambda} \exp(-t/\lambda).
\ee
Hence the distribution of $\Delta v_s$ (with $r$  fixed) is
\be
P(\Delta v_s)
= \frac{dn}{dt} \frac{dt}{d\Delta v_s}
= \frac{r}{(\Delta v_s)^2 \lambda}
  \exp\left(-\frac{r}{|\Delta v_s| \lambda}\right). 
\ee
Particle subdivision reduces
$r_\sigma$  as
$r_\sigma(l) = r_\sigma(1)/\sqrt{l}$.
Therefore,
in the large subdivision limit,
the subluminal and superluminal tails of the signal velocity distribution
scale as a power law
\be
P(\Delta v_s) \sim \left(\frac{v_0/\sqrt[4]{l}}{\Delta v_s}\right)^2,
\qquad v_0\equiv \sqrt{\frac{r}{\lambda}},
\ee
i.e., the distribution gets narrower as
$P_l(\Delta v_s) \sim P_1\left(\Delta v_s\sqrt[4]{l}\right)$.

The ``measured'' cascade distributions of the magnitude of the signal
propagation speed, $dn/dv_s$,
as defined via (\ref{Def:signal_velocity}),
and the magnitude of its transverse component, $dn/dv_{s,\perp}$,
are shown in Fig. \ref{Figure:Et_and_signal_velocity}. 
Though the distribution of $v_s$ is strongly peaked at $v_s = c$,
both super- and subluminal propagation are present. 
While increasing particle subdivision
decreases the deviations from the exact propagation speed $c$,
convergence is slow.
Even for a subdivision of 100,
$13\%$ of the collisions correspond
to a signal propagation velocity larger than $1.5c$.
One must keep in mind
that Fig. \ref{Figure:Et_and_signal_velocity} shows
the distribution of the signal propagation speed
measured over the length and time scale of a single collision.
On larger scales,
the deviation is reduced because the large scale signal velocity
is the sum of many small scale signal velocities.

In summary we demonstrated that the numerical artifacts
due to Lorentz violation and acausality are reduced by subdivision
and the cascade solution converges as $l$ increases.
In the $l\to\infty$ limit the cascade technique gives
the correct numerical solution
of the transport equation (\ref{Eq:Boltzmann_22}).
In practice, 
rather high subdivisions were found necessary to recover covariance.
We could explore convergence  up to $l=800$, 200, 150, and 100,
for $R_0=2$, 4, 6, and 8 fm with the workstations available to us.

\subsection{Scaling of the transport solutions}

Subdivision covariance (\ref{Eq:particle_subdivision})
actually implies that the transport equation has a broad dynamical range,
and the solution for any given initial condition and transport property
immediately provides the solution to a broad band
of suitably scaled initial conditions and transport properties. 
This is because solutions
for problems with $l$ times larger the initial density
$dN/d\eta d^2x_\perp$,
but with one $l$-th the reaction rate
can be mapped to the original ($l=1$) case for {\em any} $l$. 
We must use subdivision to eliminate numerical artifacts.
However, once that is achieved,
we have actually found the solution to a whole class of 
suitably rescaled problems. 

The dynamical range of the transport equation (\ref{Eq:Boltzmann_22}) 
is further increased by its covariance under coordinate rescaling
\be
f(x,\vec p) \to f'(x,\vec p) \equiv f\!\left(\frac{x}{l_x},\vec p\right),
\quad W \to W' \equiv \frac{W}{l_x} .
\label{Eq:rescale_x}
\ee
This is a simultaneous rescaling of space-time
{\em and} the  transition probability.
In addition,
there is  also a covariance under rescaling of the momenta
\bea
f(x,\vec p) &\to&
f'(x,\vec p) \equiv l_p^{-3}\, f\!\left(x,\frac{\vec p}{l_p}\right),
\nonumber\\
W(\{p_i\}) &\to& W'(\{p_i\})
\equiv l_p^2\, W\!\left(\left\{\frac{p_i}{l_p}\right\}\right) ,
\label{Eq:rescale_p}
\eea
such that the particle density is again unchanged.
This scaling also implies a rescaling of the mass
$m\to m'=m/l_p$.
Combining the  three scaling transformations,  we find
covariance of the transport theory under 
\bea
f(x,\vec p) &\to& f'(x,\vec p)
\equiv l_p^{-3} l \,f\!\left(\frac{x}{l_x},\frac{\vec p}{l_p}\right),
\nonumber\\
W(\{p_i\}) &\to& W'(\{p_i\})
\equiv \frac{l_p^2}{l_x l}\, W\!\left(\left\{\frac{p_i}{l_p}\right\}\right) .
\label{Eq:rescale_all}
\eea

In our calculation using MPC,
we  vary the physical parameters:
$\sigma$, $\mu$, $T_0$, $R_0$, $\tau_0$,
and  $n_{\eta,0} \equiv dN/d\eta d^2x_\perp|_{\tau_0}$
(the rapidity interval $\eta_{max} = 5$ was fixed).
Keeping in mind Eq. (\ref{Def:sigma})
and that
$$
n_\eta \equiv \left.\frac{dN}{dyd_{x_\perp}^2}\right|_{\tau}
= \int d^2 p_\perp d\eta \, m_t \ch(y-\eta) \tau
f(\vec x_\perp, \eta, \tau, \vec p_\perp, y) ,
$$
covariance under the transformation (\ref{Eq:rescale_all}) 
implies that once the solution for a particular choice of
these parameters is known,
then the solution is known
for any other choice of the parameters which are related to the original via
\bea
&&\sigma' = l_x^{-1} l^{-1} \sigma,
\qquad
T_0' = l_p T_0,
\qquad
R_0' = l_x R_0,
\nonumber\\
&&
n_{\eta,0}' = l_x l n_{\eta,0},
\qquad
\mu' = l_p \mu,
\qquad
\tau_0' = l_x \tau_0 .
\label{scaledpam}
\eea
Therefore,
we can scale one solution to others 
provided that
$\mu/T_0$, $R_0/\tau_0$,
and $\sigma n_{\eta,0} \sim \tau_0/\bar\lambda_{MFP}$ remain the same.
For example,
three times the density with one-third the cross section
leaves all three parameters the same,
hence the results can be obtained via scaling
without further computation.
Table \ref{Table:1} shows
sets of the three ratios that we mapped out via MPC.

\section{Freeze-Out}
\label{Section:freezeout}

\subsection{Hydrodynamic Freeze-Out Problem}
\label{SubSection:hydro_freezeout}

In Section \ref{Section:intro} we argued that hydrodynamics
cannot be valid during the complete evolution in nuclear collisions
because the assumption of local equilibrium breaks down.
Thus,
in spite of its appeal,
hydrodynamics cannot be compared with measurements 
without {\em additional} model assumptions needed
to specify when and how it breaks down.
The problem of determining those extra model assumptions
is the so-called freeze-out problem.

For application to nuclear collisions,
freeze-out  cannot be formulated as an expansion in $\lambda/L$
since by definition it occurs
when that ratio exceeds unity.
Hence, 
even the Navier-Stokes hydrodynamics
is inadequate to solve the freeze-out problem.

A common  freeze-out prescription,
which we here name ``Cooper-Frye frozen hydrodynamics'',
is to assume the validity of ideal hydrodynamics
up to a ``sharp'' 3D freeze-out hypersurface $\sigma^\mu(x)$.
Assuming that all interactions suddenly cease on that hypersurface,
the final (frozen-out) invariant differential distribution of particles
is then computed via the Cooper-Frye formula\cite{CF74}:
\be
E dN = \frac{d^3p}{(2\pi)^3} d\sigma^\mu(x) p_\mu f(x,\vec p) .
\label{Eq:Cooper-Frye}
\ee
Here $d\sigma^\mu$(x) is the normal vector to the 3D freeze-out hypersurface
at the point $x$,
while $f(x,\vp)$ is assumed to be in local equilibrium
and hence,
for classical particles,
given by Eq. (\ref{Eq:local_equil_distribution}).
While this prescription is covariant and appealingly simple,
it  suffers from several well known problems\cite{ECT97,Csernai:1997xb}:

First, because 
the hydrodynamical solutions do not contain dynamical
information needed to compute the freeze-out hypersurface,
the  assumed one is simply  an {\em ad hoc} external constraint.
It is usually  parameterized
in terms of a few physically ``reasonable'' parameters, the most common
being a  freeze-out isotherm $T(\sigma^\mu)=T_f$
or freeze-out energy density $e_f$.
It is not possible to estimate the errors introduced by such a prescription.

Second,
the Cooper-Frye formula allows
negative contributions to the measurable particle yields%
\cite{Bu96,Anderlik:1999et,MCs99}.
\OLD{%
This can be circumvented 
by a simple prescription whereby the integration measure
in Eq. (\ref{Eq:local_equil_distribution})
is replaced by $\Theta(d\sigma^\mu p_\mu)$.
It is  now known
how to satisfy energy-momentum and baryon number conservation
across the hypersurface\cite{Anderlik:1999et,ALC98}.
}
\NEW{%
This can be avoided \cite{Anderlik:1999et,MCs99,ALC98}
by choosing a non-equilibrium post freeze-out distribution
that does not have particles in the phase space domain where 
$d\sigma^\mu p_\mu < 0$.
}
However,
such a choice still relies on the existence
of a sharp 3D freeze-out surface.

Finally,
while an idealized   sharp freeze-out surface 
 may be adequate for applications to quasi-stationary macroscopic systems,
it cannot be justified in expanding mesoscopic systems
in which $L/\lambda$ is never large.
The very fact that such systems do  freeze out, 
i.e., $\lambda(\sigma)> L$,
means that  the solution to freeze-out problem
must entail {\em global} information
as the system  becomes more and more dilute.
Furthermore,
there is no way to justify the neglect of final state interactions
during freeze-out stage of the reactions
while expansion and rarefaction are causing the system to depart
from local equilibrium.

In Ref. \cite{Grassi:1996ng} 
a continuous emission hydrodynamical freeze-out model
was proposed to  overcome some of these problems.
The global information relevant to freeze-out in that model
is taken there as the Glauber escape probability
\be
P(x,p)= \exp\left( -\int_{\tau}^{\tau_{out}} d\tau^\prime \sigma v_{rel}
n(x(\tau))\right) . 
\label{Eq:kodama}
\ee
This formula reveals clearly
the highly nonlocal character of the freeze-out problem.
At any point in spacetime, $x^\mu$,
the line integral runs over the future trajectory
and therefore is exponentially sensitive
to the future evolution of the system.
This leads to a formidable self-consistency problem.
For special geometries such as  Bjorken boost invariant expansion,
a  rough  estimate of $P$ can be made
using the approximate scaling Bjorken hydrodynamic solution
\be
n(x(\tau))= \frac{\tau_0}{\tau} n(x(\tau_0))
 \; \Theta(R^2-(\vx_\perp+ \vv_\perp \tau)^2) .
\ee
Together with the Glauber straight line trajectory,
this leads to the characteristic  power law survival probability
\be
P\sim \left(\frac{\tau}{\tau_{out}}\right)^{\sigma \tau_0 n(\tau_0)} ,
\ee
that also appears, e.g., 
in the $J/\psi$ suppression problem\cite{Gavin:1988hs}.
While Eq. (\ref{Eq:kodama}) captures  essential 
global physics of freeze-out,
it is not  complete since before freeze-out the trajectories cannot be straight
if local equilibrium is maintained via  the assumed
hydrodynamic equations.
Also, in the surface
region 
where $P\sim 1/2$ neither hydrodynamics nor eikonal dynamics
applies.

The solution to the freeze-out problem in classical mechanics is given by
microscopic transport theory.
A hybrid approach that partially reaches that
end was proposed in Ref. \cite{Bass:1999tu}, which combines partonic
hydrodynamics with hadronic transport theory.
In that approach, hydrodynamics
is assumed to hold up to only some {\em intermediate} critical temperature
hypersurface, $T(\sigma_{int}^\mu)=T_c > T_f$, on which the fluid is converted
to hadrons via the Cooper-Frye formula.  Subsequent evolution of the hadronic
system is then calculated by solving the hadronic transport theory as encoded
in UrQMD.
As noted in Ref. \cite{Grassi:1996ng},
the freeze-out surface is actually a diffuse four-volume,
and in addition different hadronic species
freeze out over different four-volume domains. 
This sequential freeze-out leads to strong observable correlations
such as the mass dependence of final transverse spectra.
The main limitation (and/or advantage)
of the above hybrid model\cite{Bass:2000ib}
is the need to assume the validity of hydrodynamics
in the dense partonic phase of the collision.
It is advantageous in that possible collective effects
due to the quark-gluon confinement transition
can be explored with hydrodynamics using ``realistic'' equations of state.
It is disadvantageous in that it is far from clear
that local equilibrium is ever reached during the evolution.
Recall \cite{Gyulassy:1997ib} 
that dissipative effects on even global observables 
such as the transverse energy per unit rapidity
cannot be accurately calculated using the Navier-Stokes equations.

Despite these known complications of the freeze-out problem,
ideal hydrodynamics and Cooper-Frye freeze-out
are still commonly used to fit experimental data
using isotherm freeze-out hypersurfaces
and draw inferences about the underlying dynamics.
The consistency and significance of interpretations
based on such fits can only be assessed
by comparing detailed dynamical transport calculations
to the hydrodynamic limit (see  Section \ref{Section:results}).

\subsection{Formal Definition for Freeze-out}
\label{SubSection:formal_freezeout}

An important experimental observable
aspect of the space-time evolution of kinetic theory
is the freeze-out distribution.
In the framework of discrete parton cascade dynamics,
the  definition of the freeze-out distribution, $dN_{fo}$,
is the number of partons per $d^4x d^4p$ invariant phase space volume
with momentum $p^\mu$
that have a collision at $x^\mu$ but suffer no more collisions.
Given the trajectories,
$(\vx_a(t),\vp_a(t))$
or the world lines $x^\mu_a(\tau)$ of all partons $a$,
that distribution is given by the ensemble average
of the  space-time coordinates, 
$x_{af}^\mu\equiv (t_{af},\vx_a(t_{af}))$,
of the {\em last} collision
together with the final outgoing momentum, $\vp(t_{af}+0^+)$:
\bea
&&
f_{fo}(x,p) \equiv \frac{dN_{fo}}{d^4xd^4p}=\left\langle \sum_a \delta(t-t_{af})
\delta^3\left(\vx-\vx_{a}(t_{af})\right) \right.
\nonumber\\
&&
\qquad\qquad\qquad\qquad\qquad
\left.
\raisebox{0cm}[0.4cm][0.4cm]{$
\times
\ 
\delta^4\left(p-p_a(t_{af}+0^+)\right)
$}
 \right\rangle .
\label{nfoc}
\eea
Because the freeze-out times, $t_{af}$,
are distributed 
over a broad time interval,
$f_{fo}$ does {\em not} correspond to 
\be
f(x,p)= N \left\langle \int d\tau 
\delta^4(x-x(\tau))\delta^4(p-p(\tau)) \right\rangle
\label{fdef}
\ee
at any time or on any fixed 3-D hypersurface.
Note that $f$ measures the phase space density
of the world lines $x^\mu(\tau)$
and their four-velocities at a single point $x^\mu$.
On the other hand,
$f_{fo}$ measures the phase space density of last scattering events,
where the momentum $p$ of a particle was last changed.
Pion interferometry\cite{Padula:1990ws}
measures the Fourier transform of $f_{fo}$.
Note that  even  after integrating over the freeze-out points,
the final observed momentum spectrum, 
$\int d^4 x f_{fo}(x,p)$,
is only equal to the Cooper-Frye formula if $x_{af}^\mu$ happen to lie
on a sharp 3D hypersurface,
$\sigma^\mu(\zeta_1,\zeta_2,\zeta_3)$.
We can write the Cooper-Frye freeze-out distribution then as
\be
\frac{dN_{CF}}{d^4xd^4p}=N \int d^3{\bf \zeta} 
\delta^4(x-\sigma({\bf \zeta}))\delta^4(p-p(\sigma({\bf \zeta}))) .
\label{cffo}
\ee

As discussed in the Appendix,
we can write the (on-shell) freeze-out distribution
in terms of the solution of the Boltzmann equation as
\bea
&&E_1 \frac{dF_{fo}(x,\vec p_1)}{d^4x d^3 p_1} \equiv P_0(x,\vec p_1)
\times 
\left[\raisebox{0cm}[0.5cm][0.5cm]{$S(x, \vec p_1)$}\,\, + \right.
\nonumber\\
&&
\qquad
\left.
+\,\, 2 \int\limits_3\!\!\!\! \int\limits_4\!\!\!\! \int\limits_5
W_{34\to 15} \delta^4(p_3+p_4-p_1-p_5)
f_3 f_4\right] .
\label{Eq:freeze-out_definition}
\label{ourfo}
\eea
While neither normalized nor unique,
this expression provides at least a  {\em formal} definition
of the freeze-out distribution for the Boltzmann equation
solely in terms of  the phase space distribution $f(x,\vec p)$,
and the assumed transition probabilities $W_{ij\to kl}$.

\section{Numerical Results}
\label{Section:results}

\subsection{Kinetic versus Hydrodynamic Evolution}
\label{SubSection:evolution_comparison}

\OLD{%
In order to compare transport and hydrodynamic calculations,
}
\NEW{%
To test the ideal hydrodynamical assumptions against transport theory,
}
it is essential to eliminate as many model differences as possible.
For example, both the hydrodynamic model and the kinetic theory
should have the same degrees of freedom,
the same equation of state,
and the same initial conditions.
Equation (\ref{Eq:Boltzmann_22}) describes a gas
that in thermal equilibrium has the equation of state $e=3p$,
if the partons are massless. 
We therefore used this ideal gas equation of state
in the hydrodynamical simulations.
\OLD{%
Also,
since Eq. (\ref{Eq:Boltzmann_22}) describes Boltzmann classical particles,
we must  also use
the classical distribution (\ref{Eq:local_equil_distribution})
in the Cooper-Frye formula.
}
We also chose the transport initial condition to be in local equilibrium,
since hydrodynamics is limited to such initial conditions.

The hydrodynamic algorithm used\cite{rischdumi}
is furthermore designed for particles without a conserved charge,
i.e.,
the particle number changes as dictated by chemical equilibrium.
The algorithm solves the energy-momentum conservation equation
to obtain the energy density, pressure and flow evolution.
Then, instead of the charge conservation equation,
it exploits the relation
between density, particle mass, and temperature in chemical equilibrium
to compute the freeze-out particle distribution.
It is important to note therefore
that Eq. (\ref{Eq:Boltzmann_22}) with {\em elastic} collisions
has the same hydrodynamic limit as the hydrodynamic model 
only if the partons are massless.
This is because ideal hydrodynamics conserves entropy\cite{Bjorken83}
and for massless particles in thermal and chemical equilibrium
entropy conservation is equivalent to particle number conservation.%
\footnote{
Because in this case $s=4n$.
}
For massive particles,
we would have to compare transport to hydrodynamics
 with particle conservation.
Conversely,
we would need to supplement Eq. (\ref{Eq:Boltzmann_22})
to include inelastic channels,
such as $2\leftrightarrow 3$ in Ref. \cite{molnar:QM99},
to compare to chemically equilibrated hydrodynamics.
In the infinite rate limit
we recover the hydrodynamic model 
even though we have a fixed number of particles.
However,
when the solution is out of equilibrium (either thermal, chemical, or both),
it does make a difference
whether we include particle number changing processes or not.

\NEW{%
To test whether ideal hydrodynamics is an adequate description
of the parton transport theory (\ref{Eq:Boltzmann_22}),
we compare the evolution of the transverse energy $dE_t/dy$ at midrapidity
from the two models.
This comparison is free from any hydrodynamic freeze-out prescription
because the transverse energy
is given directly by the phase space distribution as
\bea
\left.\frac{dE_t}{dy}\right|_{\tau}
&=& \tau \int d^2 p_\perp d\eta\, d^2 x_\perp m_t \cosh(y-\eta)
\nonumber\\
&&\qquad \times \ m_t \, f(y,\vec{p_\perp}, \eta, \vec{x_\perp}, \tau),
\eea
where,
through the local equilibrium ansatz (\ref{Eq:local_equil_distribution}),
the hydrodynamic phase space evolution is determined
by the evolution of the flow velocity and local temperature
as dictated by the equations of motion (\ref{Eq:cons_laws}).
}

\NEW{%
Figure \ref{Figure:et_evolution} shows the transverse energy evolution
from transport theory and hydrodynamics,
for an initial Bjorken cylinder radius of 2 fm,
with $\tau_0 = 0.1$~fm$/c$,
$T_0=\mu=0.5$~GeV,
$n_{\eta,0}=2.6505$~fm$^{-2}$ (via scaling),
$\sigma=$ 15, and 60 mb.
(We chose $\eta_m=5$,
subdivisions 800 for 3+1D, 256 for 1+1D,
and a $100$~fm$^2$ transverse area for the 1+1D evolution.)
}

\NEW{%
The transverse energy decreases much faster
from ideal hydrodynamics than from kinetic theory,
both in 1+1D and 3+1D,
showing that hydrodynamics does more work than the cascade.
This is due to the different phase space evolution in the two models.
The early discrepancy, even for cross sections as extreme as 15 or 60mb,
indicates
that either the transport evolution gets very quickly out of equilibrium,
or the {\em initial} evolution is close to equilibrium
but the energy-momentum tensor is not ideal.
Note that even if the latter is true,
it does not necessarily mean
that this initial, locally equilibrated, nonideal dynamics
can be described by the Navier-Stokes equations.
}

\NEW{%
The above conclusion holds for any initial system size
larger than 2fm as well.
Since the 1+1D curves correspond to the infinite transverse size limit,
the hydrodynamic and transport evolutions for initial sizes larger than 2fm
will lie between the 2fm and the 1+1D curves
for hydrodynamics and for transport theory, respectively.
Because these two regions do not overlap,
the discrepancy between ideal hydrodynamics and transport theory
will not disappear with increasing system size.
}

\subsection{Kinetic vs Hydrodynamic Freeze-out Results} 
\label{SubSection:freezeout_comparison}

\NEW{%
In the previous Subsection
we showed that parton kinetic theory does not reduce to ideal hydrodynamics
for initial conditions at RHIC.
Thus,
the final observables from the two models can be similar
{\em only if} the hydrodynamic freeze-out prescription helps mimic
the observables from the nonequilibrium transport evolution.
}

\NEW{%
Here we test whether one can reproduce the transport observables
by a suitable choice of the hydrodynamic freeze-out parameters.
We chose the widely-used Cooper-Frye freeze-out
prescription (\ref{Eq:Cooper-Frye}) with
isotherm freeze-out surfaces,
despite all known problems discussed in Section \ref{Section:freezeout}.
Hence,
our only adjustable parameter is the freeze-out temperature.
Since Eq. (\ref{Eq:Boltzmann_22}) describes Boltzmann classical particles,
we must use
the classical distribution (\ref{Eq:local_equil_distribution})
in the Cooper-Frye formula.
}

\subsubsection{Coordinate space evolution}
\label{SubSubSection:coordinate_space}

\begin{figure}[hp]
\center
\leavevmode
\epsfysize=14cm
\epsfbox{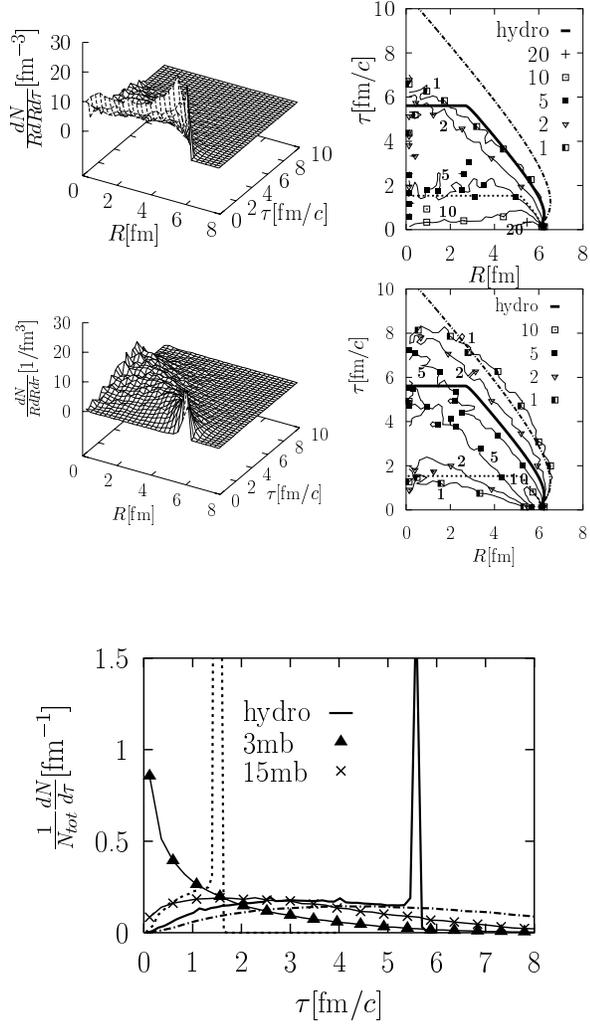}
\caption{
\footnotesize
The left column shows
the transverse coordinate and proper time distribution $\frac{dN}{RdRd\tau}$,
of freeze-out coordinates.
Top row corresponds to $\sigma=3$ mb  and middle row to 15 mb.
The initial Bjorken cylinder radius is $R_0=6$ fm in both cases.
The right column shows  contour plots corresponding to the left column.
The thick lines show  Cooper-Frye isotherms:
$T_f=100$ MeV (dashed-dotted line), 130 MeV (thick solid line),
and 200 MeV (dotted line).
The bottom figure compares  the proper time freeze-out distribution,
$\frac{dN}{d\tau}$, 
for the different cases. 
}
\label{Figure:3}
\label{Figure:fo_spacetime_6fm}
\end{figure}

\begin{figure}[hp]
\center
\leavevmode
\epsfysize=14cm
\epsfbox{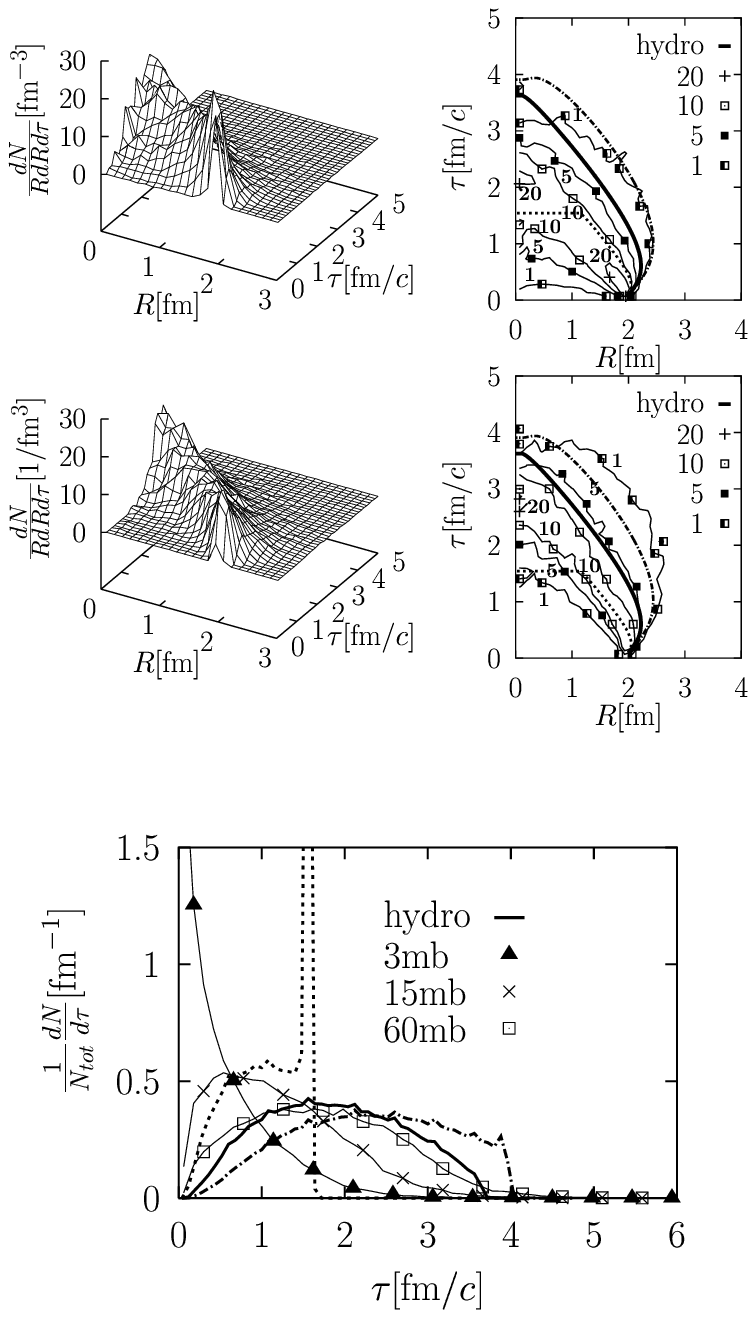}
\caption{
\footnotesize
The left column shows
the transverse coordinate and proper time distribution
$\frac{dN}{RdRd\tau}$, of freeze-out coordinates.
Top row corresponds to $\sigma=15$ mb  and middle row to 60mb.
The initial Bjorken cylinder radius is $R_0=2$ fm in both cases
in contrast to Fig. \ref{Figure:fo_spacetime_6fm}, where $R_0=6$ fm.
The right column shows  contour plots corresponding to the left column.
The thick lines show  Cooper-Frye isotherms:
$T_f=100$ MeV (dashed-dotted line), 130 MeV (thick solid line),
and 200 MeV (dotted line).
The bottom figure compares  the proper time freeze-out distribution,
$\frac{dN}{d\tau}$, 
for the different cases. 
}
\label{Figure:5}
\label{Figure:fo_spacetime_2fm}
\end{figure}

Freeze-out distributions in space-time from MPC
are shown in Figs. \ref{Figure:3} and \ref{Figure:5}.
Due to the assumed cylindrical symmetry and longitudinal boost invariance,
that distribution is only a function of $\tau$ and $R$.

Figures \ref{Figure:fo_spacetime_6fm}, \ref{Figure:fo_spacetime_2fm}
show the freeze-out distribution
for initial radii 6 fm and 2 fm, respectively,
with 
$\tau_0=0.1\ {\rm fm}/c$,
$T_0=\mu = 0.5$ GeV,
$n_{\eta,0}=2.6505$~fm$^{-2}$ (via scaling),
$\sigma=$ 3, 15, and 60 mb.
For comparison,
three different freeze-out isotherms are also shown
from solution of Cooper-Frye frozen ideal hydrodynamics.
(We chose $\eta_m=5$,
subdivisions 800 for 2 fm, and 150 for 6 fm.)

Unlike the sharp hydrodynamic freeze-out surface,
the freeze-out distribution from the cascade is a broad {\em wedge}.
Particles originate from a hypervolume in space-time,
rather than from a hypersurface.
In the top left plot (3 mb, 6 fm) in Fig. \ref{Figure:fo_spacetime_6fm},
the wedge moved down to $\tau=\tau_0$,
which is a general feature for very low reaction rates.
In the limit of a vanishing reaction rate,
all particles freeze out from $\tau=\tau_0$.

Figures \ref{Figure:fo_spacetime_6fm}, \ref{Figure:fo_spacetime_2fm} show
that particles freeze out later with increasing microscopic rates
as expected.
The maximum of the wedge
moves outward with increasing rates,
hence no freeze-out temperature can be universal.
If we tune the freeze-out temperature
to get as close as possible to the cascade freeze-out distribution,
the freeze-out temperature will depend on the reaction rate.

Thus, the remarkable agreement
seen in the bottom figure of Fig. \ref{Figure:fo_spacetime_2fm}
between 
\NEW{%
Cooper-Frye frozen
}
ideal hydrodynamics with a 130 MeV freeze-out temperature
and the cascade with $\sigma=60$ mb
is a mere coincidence;
higher rates would lead to a later freeze-out. 
For very high reaction rates,
the 130-MeV hypersurface {\em from the cascade}
would be very close to that from hydrodynamics
because the hydrodynamic evolution
is the infinite reaction rate limit of the cascade evolution.
But that does not mean that the freeze-out distributions are the same.
On the contrary,
if hydrodynamics and the cascade are close to each other at $T=130$ MeV
then we have no justification
to stop the
\NEW{%
hydrodynamic
}
evolution and freeze out with Eq. (\ref{Eq:Cooper-Frye})
because we are still in equilibrium
and particles will certainly collide in the future,
i.e., they have not yet frozen out.

It is {\em not} possible to tune the
\NEW{%
Cooper-Frye
}
freeze-out temperature to reproduce
the cascade freeze-out distribution.
Though the contour plots in Figs.
\ref{Figure:fo_spacetime_6fm}, \ref{Figure:fo_spacetime_2fm}
suggest that such a tuning
can get the hydrodynamic freeze-out {\em curve}
close to the {\em ridge of the wedge} of the cascade freeze-out distribution,
that is not enough.
As the $dN/d\tau$ distributions show,
the resulting hydrodynamic freeze-out distribution
is {\em not} close to the cascade distribution
because one has to reproduce not only the curve given
by the ridge of the wedge
but {\em also} the exact distribution along this curve.

If the freeze-out temperature is high enough to yield
a freeze-out surface with a timelike portion,
we get unphysical spikes in the freeze-out distribution
that are not present in the cascade calculations.
This can be seen in Fig. \ref{Figure:fo_spacetime_2fm} for $T_f=200$ MeV,
and in Fig. \ref{Figure:fo_spacetime_6fm} for $T_f=130$ and 200 MeV.
For example, for $R_0=6$ fm with $T_f=130$ MeV, 
\NEW{%
Cooper-Frye frozen
}
hydrodynamics produces most particles at around $\tau=5.6$~fm$/c$.
Cooper-Frye frozen hydrodynamics produces most particles
at around $\tau=5.6$~fm$/c$.
This is because the inside of the cylinder follows a 1D Bjorken evolution
with $T(\tau)=T_0(\tau_0/\tau)^{1/3}$
until the rarefaction wave from the boundary arrives.
The rarefaction wave travels with a speed $c_s=1/\sqrt{3}$.
If the system is large enough,
most of the system reaches the freeze-out temperature
{\em before} the rarefaction wave arrives, i.e.,
during the 1+1D Bjorken evolution.
With our parameters $T_0=0.5$ GeV,
$T_{fo}=130$ MeV,
and $\tau_0=0.1\ {\rm fm}/c$,
this gives a freeze-out for the {\em inside} of the cylinder
at $\tau_{fo}=5.6\ {\rm fm}/c$,
which is in complete disagreement with our transport theory solutions.
Furthermore,
it does not correspond to the infinite reaction rate limit either
because in that case  particles freeze out very late.

Hence the peaks in $dN/d\tau$ at $\tau=5.6$~fm$/c$ ($T_f=130$ MeV)
and $\tau=1.6$~fm$/c$ ($T_f=200$ MeV)
are a clear consequence
of the arbitrary freeze-out prescription
using  Eq. (\ref{Eq:Cooper-Frye}) with isotherm freeze-out hypersurfaces.
Smearing the peaks out around their maxima does not help either
because that does not change the location of the peaks,
while the maximum from the cascade moves outward
with increasing reaction rates.

\subsubsection{Momentum space}
\label{SubSubSection:momentum_space}

\begin{figure}[hp]
\center
\leavevmode
\epsfysize=11cm
\epsfbox{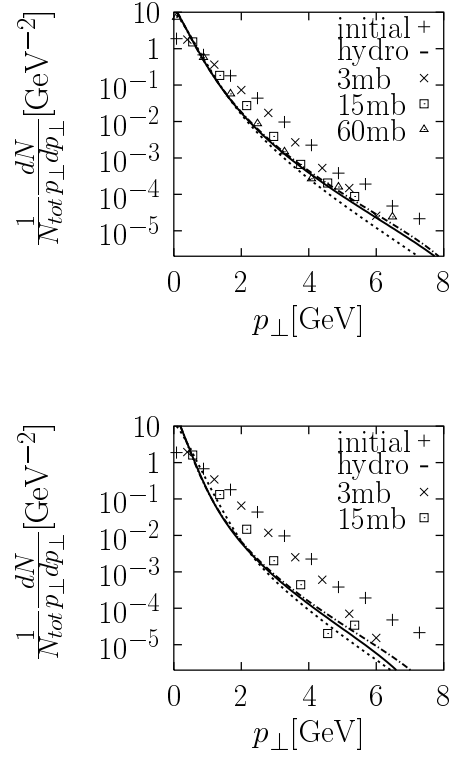}
\caption{
\footnotesize
The top figure shows the freeze-out $p_\perp$-distributions
from an initial radius $R_0=2$ fm
for the cascade with $\sigma=3$, 15, 60 mb,
and for ideal hydrodynamics with Cooper-Frye freeze-out
with freeze-out temperatures $T_f=$100 MeV (dashed-dotted line),
130 MeV (thick solid line), and 200 MeV (dotted line).
The initial $p_\perp$-distribution is shown using pluses.
The bottom figure shows the same but for an initial radius $R_0=6$ fm
with cross sections 3 and 15 mb.
}
\label{Figure:6}
\label{Figure:fo_pt}
\end{figure}

The freeze-out distribution in momentum space
is shown in Figs. \ref{Figure:2} and \ref{Figure:6}.
Figure \ref{Figure:fo_pt} shows the freeze-out $p_\perp$-distribution
for initial radii 2 fm and 6 fm,
cascade cross sections 3, 15, and 60 mb
compared to ideal hydrodynamics with a Cooper-Frye freeze-out
at temperatures $T_f=100$, 130, and 200 MeV.
As the reaction rate increases, the small $p_\perp$-slopes rise
as the system cools due to longitudinal work.
The $p_\perp$-distribution seems to 
approach that of
\NEW{%
Cooper-Frye frozen
}
hydrodynamics.
However, this is only an illusion
on a low-resolution logarithmic plot.
Figure \ref{Figure:fo_pt_ratio},
where we plotted the final $p_\perp$-spectra divided by
the initial $T_0=500$ MeV thermal one,
shows that there {\em is} a large,
up to a factor of ten
difference at both low ($<0.5$ GeV) and high $p_\perp$ ($>2$ GeV),
depending on the microscopic rates.
For all the cases studied,
\NEW{%
Cooper-Frye frozen
}
hydrodynamics has more low-$p_\perp$ particles but fewer high-$p_\perp$ ones
than the cascade.
This is not necessarily a general feature
because the assumed hydrodynamic freeze-out temperature
is an arbitrary number.
A later freeze-out (lower temperature) gives a larger slope,
an earlier freeze-out (higher temperature) gives a smaller one.

It is also striking that one would need rather high,
$T_f\sim 300-450$ MeV freeze-out temperatures
to get closer to the cascade $p_\perp$-spectra.
We conclude that  it is not possible to reproduce
{\em both} the space-time and the momentum space
transport theory freeze-out distributions
\OLD{%
with Cooper-Frye freeze-out prescription.
}
\NEW{%
using ideal hydrodynamics with
the isotherm Cooper-Frye freeze-out prescription.
Either one needs to treat hydrodynamic freeze-out more accurately
than the Cooper-Frye prescription,
or one needs to use full-scale transport theory instead of ideal hydrodynamics.
The present work is a step in the latter direction,
while Refs. [21-23] are important steps in the former direction
looking for a simplification of the full transport theoretical problem
that, hopefully, will still be applicable to a wide class of situations.
}

\section{Outlook}
\label{Section:outlook}

There are many open problems in the development
of covariant transport theory. The most urgent need
is to develop practical convergent algorithms
to  incorporate inelastic $2\leftrightarrow 3$ processes
to allow studies of chemical equilibration.
Preliminary work in Ref. \cite{molnar:QM99} indicated
a rather slow convergence towards Lorentz covariance
with particle subdivision.
Unlike the $l^{-1/2}$ convergence in $2\to2$,
a much slower $\sim l^{-1/5}$ convergence
is expected in $2\leftrightarrow 3$ processes
even when nonlocal formation physics ($\Delta t > \hbar/\Delta E$)
is neglected.

Also, 
we note that all results in this paper pertain
to homogeneous initial conditions.
In Ref. \cite{turb},
it was shown that jets induce large nonstatistical local fluctuations
that may evolve in a turbulent manner.
A transport study of the evolution from such inhomogeneous initial conditions
would be useful to compare to the known hydrodynamic solutions.

\section{Acknowledgments}

We are grateful to Yang Pang and Bin Zhang
for extensive discussions on transport theory 
and numerical cascade algorithms
and their contributions to the OSCAR effort.
We are also grateful to Adrian Dumitru and Dirk Rischke
for discussions on hydrodynamics and use of their codes.
We acknowledge useful discussions
with L\'aszl\'o P. Csernai and George Bertsch
on hydrodynamics and freeze-out,
and the Parallel Distributed Systems Facility
at the National Energy Research Scientific Computing Center
for providing computing resources.

This work was supported by the Director, Office of Energy Research,
Division of Nuclear Physics of the Office of High Energy and Nuclear Physics
of the U.S. Department of Energy under contract No. DE-FG-02-93ER-40764.

\appendix

\section*{Formal definition for freeze-out}

Unlike in the cascade solution
where the freeze-out distribution is trivially defined by Eq. (\ref{nfoc}), 
in the Boltzmann equation $f$ changes continuously
and no discrete final collisions can be identified.
In this appendix we propose a generalization of Eq. (\ref{Eq:kodama})
which is independent of the discrete numerical cascade picture.
We motivate here  a {\em formal} definition, Eq. (\ref{ourfo}),
of the freeze-out distribution
using solely $f(x,\vec p)$ and  $W_{ij\to kl}$.

Following the notion of the ``last collision'',
one can first compute the probability
that a particle  starting at a coordinate $x_1^\mu$ with momentum $\vec p_1$
does {\em not} collide any further.
The collision rate is given by
\bea
\Gamma_{coll}&\equiv&\frac{dN_{coll}}{d^4 x}(x,\vec p_1,\vec p_2)
\nonumber\\
&=&
f_1(x,\vec p_1) f_2(x,\vec p_2) \sigma(p_1, p_2) v_{12} d^3 p_1 d^3 p_2,
\label{Def:coll_rate}
\eea
where
the relative velocity and the total cross section
are given with the Lorentz scalar 
\be
t_{12} \equiv \sqrt{(p_1^\mu {p_2}_\mu)^2 - m_1^2 m_2^2}
\ee
as
\bea
&&\qquad\qquad\qquad\qquad v_{12} = \frac{t_{12}}{E_1 E_2},
\nonumber\\
&&\sigma(p_1, p_2)
=
\frac{1}{t_{12}}
\int\limits_3 \!\!\!\! \int\limits_4 \! W_{12\to 34} \delta^4(p_1+p_2-p_3-p_4)
.
\label{Def:sigma}
\eea

A free pointlike particle has the phase space distribution
\be
f_1(x,p) = \int_0^\infty d\tau \delta^4(x-x_1-u_1\tau)
\delta^4(p-p_1),
\ee
where for an on-shell particle
$$
 \delta\left(p^0 - \sqrt{\vec p^2+ m^2}\right) = \delta(p^2 - m^2)
 2 \sqrt{m^2 + \vec p^2} \Theta(p_0),
$$
i.e.,
$$
\delta^4(p - p_1) = \delta^3(\vec p - \vec p_1) \delta(p^2-m^2)
2 \sqrt{m^2 + \vec p_1^2} \Theta(p_0)
$$
and thus%
\footnote{
Recall, the on-shell phase space distribution is defined via
\be
f(x, p) \equiv 2m \,\delta(p^2 - m^2) \Theta(p_0) f(x, \vec p).
\label{Eq:f_onshell}
\ee
}
\bea 
f_1(x,\vec p) &=& \int_0^\infty d\tau\, \delta^4(x-x_1-u_1\tau)
\nonumber\\
&&\qquad\quad
\times\,\,
 \delta^3(\vec p-\vec p_1)
\frac{\sqrt{m^2+\vec p_1^2}}{m}.
\eea
Plugging this result into Eq. (\ref{Def:coll_rate}),
the probability
that a free particle will not have any further collisions is\footnote{
This can be derived assuming that subsequent collisions are uncorrelated
(just like the similar formula
for the inhomogeneous Poisson distribution).
}
\bea
&&P_0(x_1,\vec p_1) = \exp\left(-\int \Gamma_{coll}
d^3 p_1 d^3p_2 d^4x \right)
\nonumber\\
&&\ \ = 
\exp\left( -\int \frac{d\tau d^3 p_2}{E_2 m} f_2(x_1+u_1\tau, \vec p_2) \sigma(p_1, p_2) t_{12} \right).
\label{Eq:P0_definition}
\eea

Now we can write the freeze-out distribution
as the number of particles having a collision at $x^\mu$
with {\em outgoing} momentum $\vec p_1$
times the probability that these particles do not collide any further, i.e.,
\bea
&&E_1 \frac{dF_{fo}^{coll}(x,\vec p_1)}{d^4x d^3 p_1} \equiv
P_0(x,\vec p_1)
\nonumber\\
&&
\qquad\qquad
\times \,\,
2 \int\limits_3\!\!\!\! \int\limits_4\!\!\!\! \int\limits_5\!
W_{34\to 15} \delta^4(p_3+p_4 - p_1 - p_5) f_3 f_4 .
\label{Eq:freeze-out_definition_coll}
\eea

This definition does not include those particles that are formed
but suffer no collisions afterward.
Their contribution is
\be 
E_1 \frac{dF_{fo}^{form}(x,\vec p_1)}{d^4x d^3p_1} \equiv
S(x, \vec p_1) P_0(x,\vec p_1).
\label{Eq:freeze-out_definition_form}
\ee
Hence the final freeze-out distribution is given by
Eq. (\ref{ourfo}).

The definition (\ref{Eq:freeze-out_definition}) should be regarded
only one measure of the freeze-out distribution
because it has several a shortcomings.
The probabilities summed are not probabilities for disjoint events.
One should {\em exclude} the volume in space time
given by all the linear paths of the already frozen-out particles.
This requires knowledge of multiparticle correlations
beyond the scope of the Bolztmann equation.
As long as those excluded volume effects are small,
Eq. (\ref{ourfo}) is adequate.
A clear problem with the present formal definition
is that particle number and momentum are not conserved by it
as is automatic in (\ref{nfoc}).
It is interesting to contrast on the other hand,
the trivial way that the cascade solves this problem
through Eq. (\ref{nfoc}).
In cascade,
the $N$-body correlations are automatically calculated
and freeze-out is easily defined conserving number and  total four-momentum.
The continuum limit is thus subtle.
Our numerical results  define that continuum limit
as the limit of infinite subdivisions using the cascade technique.


%
%

\newpage

\NEW{%
\begin{figure}[hp]
\center
\leavevmode
\hskip 0cm
\epsfysize=8cm
\epsfbox{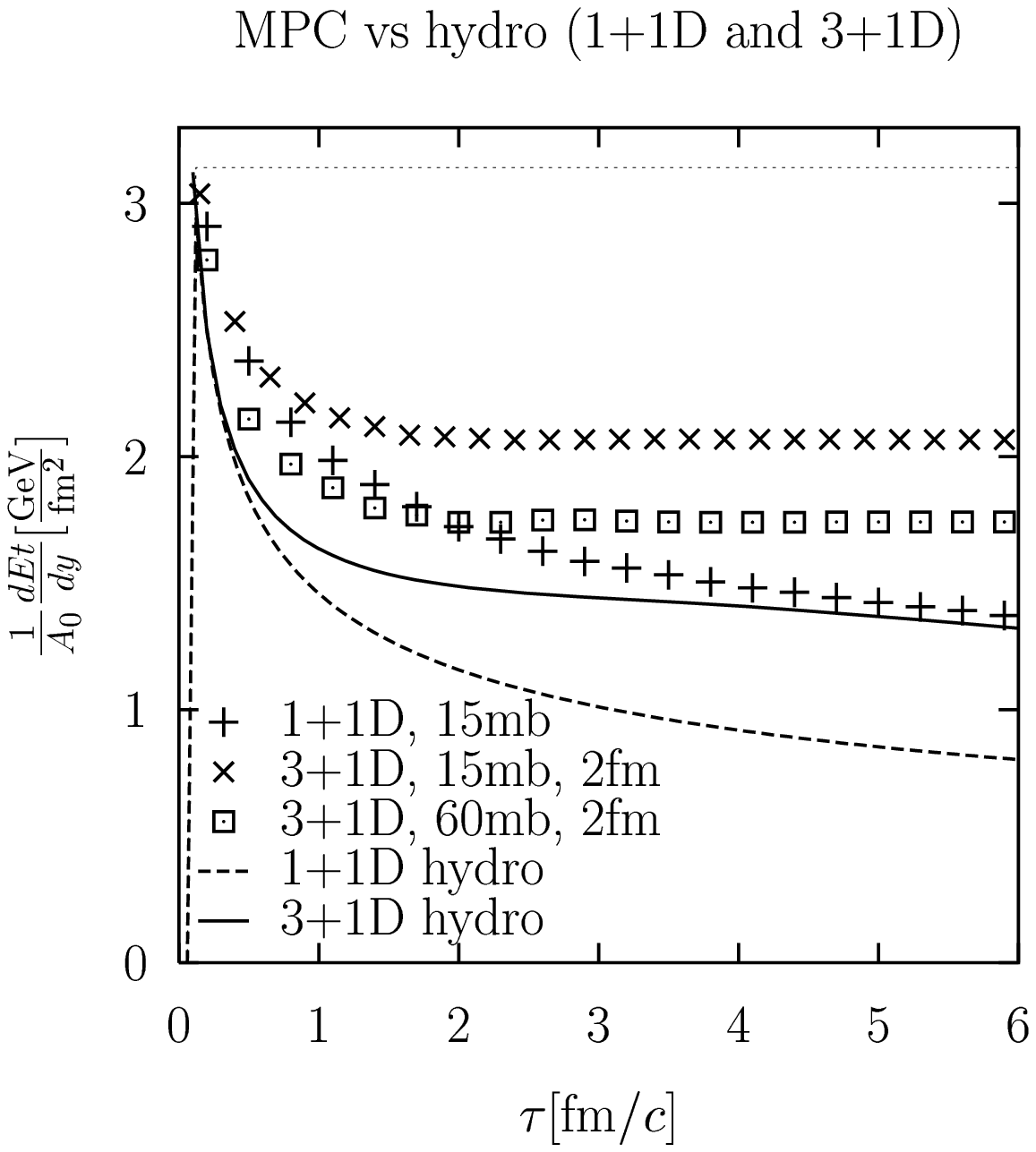}
\caption{
\footnotesize
This figure shows the evolution of the transverse energy $dE_t/dy$
at midrapidity,
normalized by the initial transverse area,
from kinetic theory and from hydrodynamics
both for 1+1 (transverse periodic) and 3+1 dimensions.
The initial distribution was a Bjorken cylinder with a radius
$R_0=2$~fm at proper time $\tau_0=0.1$~fm$/c$
in local thermal and chemical equilibrum at $T_0=500$ MeV.
The cross sections were $\sigma=15$ and 60 mb,
with the cutoff $\mu = 0.5$~GeV.
Note,
that the hydrodynamical results are {\em free from
any arbitrary freeze-out prescription}
because they depend only on the evolution of the phase space distribution
ansatz (\ref{Eq:local_equil_distribution})
as dictated by the equations of motion (\ref{Eq:cons_laws}).
The difference between the hydrodynamical and the kinetic theory results,
even for a large, 60 mb cross section,
indicates that ideal hydrodynamics is not applicable
for the evolution.
}
\label{Figure:1}
\label{Figure:et_evolution}
\end{figure}
}


\newpage

\newpage

%
%

\newpage

\begin{table}[]
\begin{center}
\begin{tabular}{|c|c|c||c|c|c||c|c|c|@{}}
\hline
$\mu/T_0$ & $R_0/\tau_0$ & $\sigma n_{\eta,0}$ &
$\mu/T_0$ & $R_0/\tau_0$ & $\sigma n_{\eta,0}$ &
$\mu/T_0$ & $R_0/\tau_0$ & $\sigma n_{\eta,0}$ \\
\hline
\hline
0 & 20 & 4 &
1 & 6 & 4 &
1 & 40 & 8 \\
\hline
0 & 40 & 4 &
1 & 8 & 4 &
1 & 40 & 16 \\
\hline
0.2 & 20 & 4 &
1 & 20 & 0.8 &
1 & 60 & 0.8 \\
\hline
0.2 & 40 & 4 &
1 & 20 & 4 &
1 & 60 & 4 \\
\hline
1 & 2 & 0.8 &
1 & 20 & 8 &
1 & 80 & 0.8 \\
\hline
1 & 2 & 4 &
1 & 20 & 16 &
1 & 80 & 4 \\
\hline
1 & 2 & 16 &
1 & 40 & 0.8 &
& & \\
\hline
1 & 4 & 4 &
1 & 40 & 4 &
& & \\
\hline
\end{tabular}

\end{center}
\caption[1]{
\footnotesize
Solutions for the above sets of parameters
were computed via MPC for the present study.
}
\label{Table:1}
\end{table}

\end{document}